\documentclass[a4]{article}

\usepackage{emulateapj}
\usepackage{apjfonts}
\usepackage{epsf}

\lefthead{Swaters, Madore, Trewhella}
\righthead{High resolution rotation curves of LSB galaxies}

\newcommand{\tskip}{\tablevspace{3pt}}
\newcommand{\HI}{\ion{H}{1}}
\newcommand{\mlstar}{$\Upsilon_{\!\!*}$}
 1

\def\ml{\ifmmode\Upsilon_{\!\!*}\else$\Upsilon_{\!\!*}$\fi}

\begin{document}

\submitted{Accepted for publication in ApJ Letters}

\title{High Resolution Rotation Curves of Low Surface Brightness Galaxies}
\author{R.A. Swaters\altaffilmark{1},
B.F. Madore\altaffilmark{2},
M. Trewhella\altaffilmark{3}}

\altaffiltext{1}{Kapteyn Astronomical Institute, University of
  Groningen, The Netherlands and Department of
  Terrestrial Magnetism, Carnegie Institution of Washington,
  Washington DC 20015, USA}
\altaffiltext{2}{Observatories of the Carnegie Institution of Washington and
  NASA/IPAC Extragalactic Database, California Institute of
  Technology, Pasadena CA 91125, USA}
\altaffiltext{3}{Infrared Processing and Analysis Center, California Institute
  of Technology, Pasadena CA 91125, USA}

\begin{abstract}

High resolution H$\alpha$ rotation curves are presented for five low
surface brightness galaxies.  These H$\alpha$ rotation curves have
shapes different from those previously derived from \HI\ observations,
probably because of the higher spatial resolution of the H$\alpha$
observations.  The H$\alpha$ rotation curves rise more steeply in the
inner parts than the \HI\ rotation curves and reach a flat part beyond
about two disk scale lengths.  With radii expressed in optical disk
scale lengths, the rotation curves of the low surface brightness
galaxies presented here and those of HSB galaxies have almost identical
shapes.  Mass modeling shows that the contribution of the stellar
component to the rotation curves may be scaled to explain most of the
inner parts of the rotation curves, albeit with high stellar
mass-to-light ratios.  On the other hand, well fitting mass models can
also be obtained with lower contributions of the stellar disk.  These
observations suggest that the luminous mass density and the total mass
density are coupled in the inner parts of these galaxies. 

\end{abstract}


\section{Introduction}

{\setlength{\baselineskip}{1.04\baselineskip}

The rotation curves of high surface brightness (HSB) spiral galaxies
rise fairly steeply to reach an extended, approximately flat part, well
within the optical disk (Bosma 1978, 1981a,b; Rubin, Ford, \& Thonnard
1978, 1980).  The discovery that the rotation curves of these galaxies
are more or less flat out to one or two Holmberg radii has been one of
the key pieces of evidence for the existence of dark matter outside the
optical disk (see also van Albada et al.\ 1985).  Within the optical
disk, the observed rotation curves can in most cases be explained by the
stellar components alone (Kalnajs 1983; Kent 1986).

The rotation curves of so-called low surface brightness (LSB) galaxies
have been studied only recently (de Blok, McGaugh, \& van der Hulst
1996, hereafter BMH; see also Pickering et al.\ 1997).  These rotation
curves, derived from \HI\ observations, were found to rise more slowly
than those of HSB galaxies of the same luminosity, if the radii are
measured in kpc.  At the outermost measured point, they were often still
rising.  McGaugh \& de Blok (1998) noted that, with radii expressed in
disk scale lengths, the rotation curve shapes of LSB and HSB galaxies
become more similar, but not necessarily identical.  Based on mass
modeling of these \HI\ rotation curves, de Blok \& McGaugh (1997,
hereafter BM) concluded that LSB galaxies were dominated by dark matter
and that the contribution of the stellar disk to the rotation curve,
even if scaled to its maximum possible value, could not explain the
observed rotation curve in the inner parts. 

The \HI\ rotation curves of LSB galaxies have received a great deal of
attention, because they provide additional constraints on theories of
galaxy formation and evolution, and dark halo structure (e.g.,
Dalcanton, Spergel, \& Summers 1997; Mihos, McGaugh, \& de Blok 1997;
Hernandez \& Gilmore 1998; Kravtsov et al.\ 1998; McGaugh \& de Blok
1998).  Unfortunately, most of the galaxies studied in BMH and BM are
only poorly resolved, making their results sensitive to the effects of
beam smearing.  For five of the galaxies in their sample, high
resolution H$\alpha$ rotation curves, which have been obtained in order
to eliminate beam smearing and to investigate the rotation curve shapes
in the inner parts of LSB galaxies, are presented in this Letter. 
}

\section{Sample, observations and data reduction}

{\setlength{\baselineskip}{1.04\baselineskip}

The galaxies presented here were selected from the sample of LSB
galaxies of BMH.  Only galaxies were chosen that satisfied the criteria
given in BM to define their useful rotation curves.  An overview of the
properties of the galaxies is given in Table~\ref{tabprops}, which lists
the name of the galaxy (1), the adopted distance in Mpc, for $H_0=75$ km
s$^{-1}$ Mpc$^{-1}$ (2), the central surface brightness in mag
arcsec$^{-2}$ (3), the disk scale length in kpc (4), the inclination
angle (5), the position angle (6), the absolute magnitude (7) and the
systemic velocity (8).

\setcounter{footnote}{3} The observations were carried out at Palomar
Observatory with the $200''$ Hale telescope\footnote{The Palomar
  $200''$ telescope is operated in a joint agreement among the
  California Institute of Technology, the Jet Propulsion Laboratory
  and Cornell University}, on November 20 1998.  The FWHM velocity
resolution was 54 km s$^{-1}$, the pixel size in the spatial direction
was $0.5''$.  Each galaxy spectrum consisted of a single 1800s
exposure.  The slit was oriented along the major axis, at the position
angle derived by BMH (see Table~\ref{tabprops}).  Despite their low
surface brightnesses, all galaxies showed up clearly on the
slit-viewing monitor.  The slit could therefore accurately be aligned
with the center of the galaxy by eye.  The data were reduced using
standard procedures in {\sc iraf} and the resulting H$\alpha$
position-velocity diagrams are presented in Fig.~\ref{figure1}.
}

\vbox{
\begin{center}
\epsfxsize=\hsize
\epsfbox{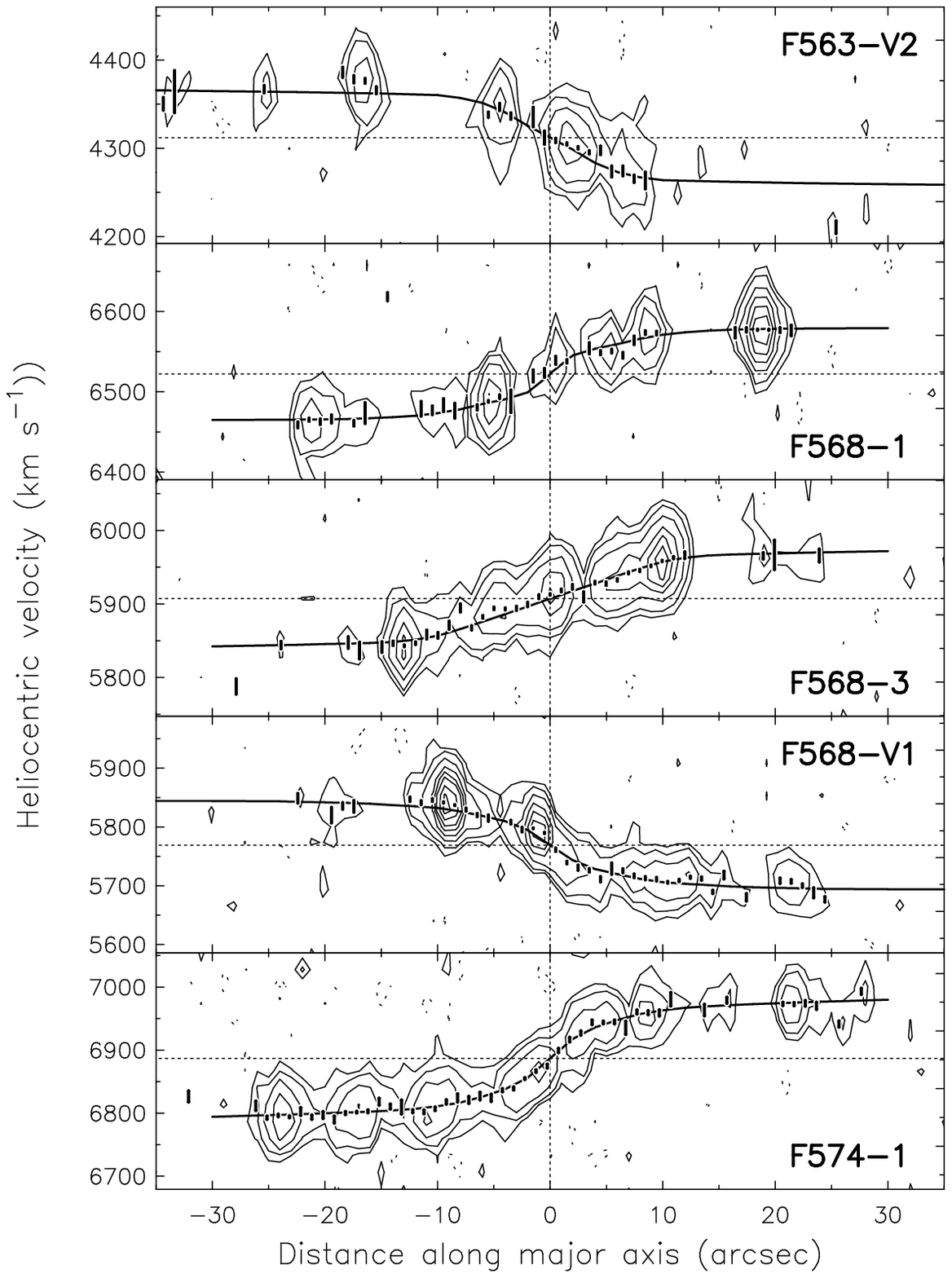}
\vskip-8pt
\figcaption{H$\alpha$ position-velocity diagrams for the five LSB
  galaxies, binned to $1''$. Contour levels are at -2, 2, 4, 8, 16,
  24, 32 times the r.m.s.\ noise.  The dots with errorbars give the
  radial velocities with the formal errors as derived from Gaussian
  fits to the velocity profiles. The solid lines represent the
  rotation curves, derived as described in section 3. The vertical
  dotted line indicates the galaxy center, the horizontal dotted lines
  denotes the heliocentric systemic velocities.\label{figure1}}
\vskip-24pt
\end{center}}

\begin{center}
{\sc Table 1\\ \smallskip Properties of the galaxies$^a$ \smallskip}
\footnotesize
\setlength{\tabcolsep}{4pt}
\begin{tabular}{lccccccc}  
\tskip \tableline
\tableline \tskip
\hfil Name & $D$ & $\mu_0^B$ & $h$ & $i$ & P.A. & $M_B$ & v$_{sys}^b$ \nl
& (Mpc) & (mag/$\prime\prime^{-2}$) & (kpc) & ($^\circ$) & ($^\circ$) &
(mag) & (km s$^{-1}$) \nl
\hfil(1) & (2) & (3) & (4) & (5) & (6) & (7) & (8) \nl
\tableline \tskip
F563-V2 & 61 & 22.1 & 2.1 & 29 & 148 & $-18.2$ & $4310\pm 4$ \\
F568-1  & 85 & 23.8 & 5.3 & 26 &  13 & $-18.1$ & $6524\pm 6$ \\
F568-3  & 77 & 23.1 & 4.0 & 40 & 169 & $-18.3$ & $5911\pm 3$ \\
F568-V1 & 80 & 23.3 & 3.2 & 40 & 136 & $-17.9$ & $5769\pm 7$ \\
F574-1  & 96 & 23.3 & 4.3 & 65 & 90$^c$ & $-18.4$ & $6889\pm 6$ \\
\tableline
\end{tabular}
\end{center} 
\vskip-6pt   
\centerline{\vbox{\hsize=8cm \footnotesize $^a$ Data from de Blok et
    al.\ (1996) and de Blok \& McGaugh (1997).}}
\vskip-2pt
\centerline{\vbox{\hsize=8cm \footnotesize $^b$ This Letter.}}
\vskip-2pt
\centerline{\vbox{\hsize=8cm \footnotesize $^c$ A P.A. of $90^\circ$
      was used, derived from the optical image in BMH.}}
\vskip20pt

\section{The high resolution rotation curves}

To derive the rotation curves, we started by making Gaussian fits to the
line profiles at each position along the major axis to obtain the radial
velocities.  These fits and their errors are overlayed on the H$\alpha$
position-velocity diagram in Fig.~\ref{figure1}.  The positions of the
galaxy centers were determined from the peak of the continuum light
along the slit.  All galaxies were suffi-\break

\vbox{\begin{center}
\epsfxsize=\hsize
\epsfbox{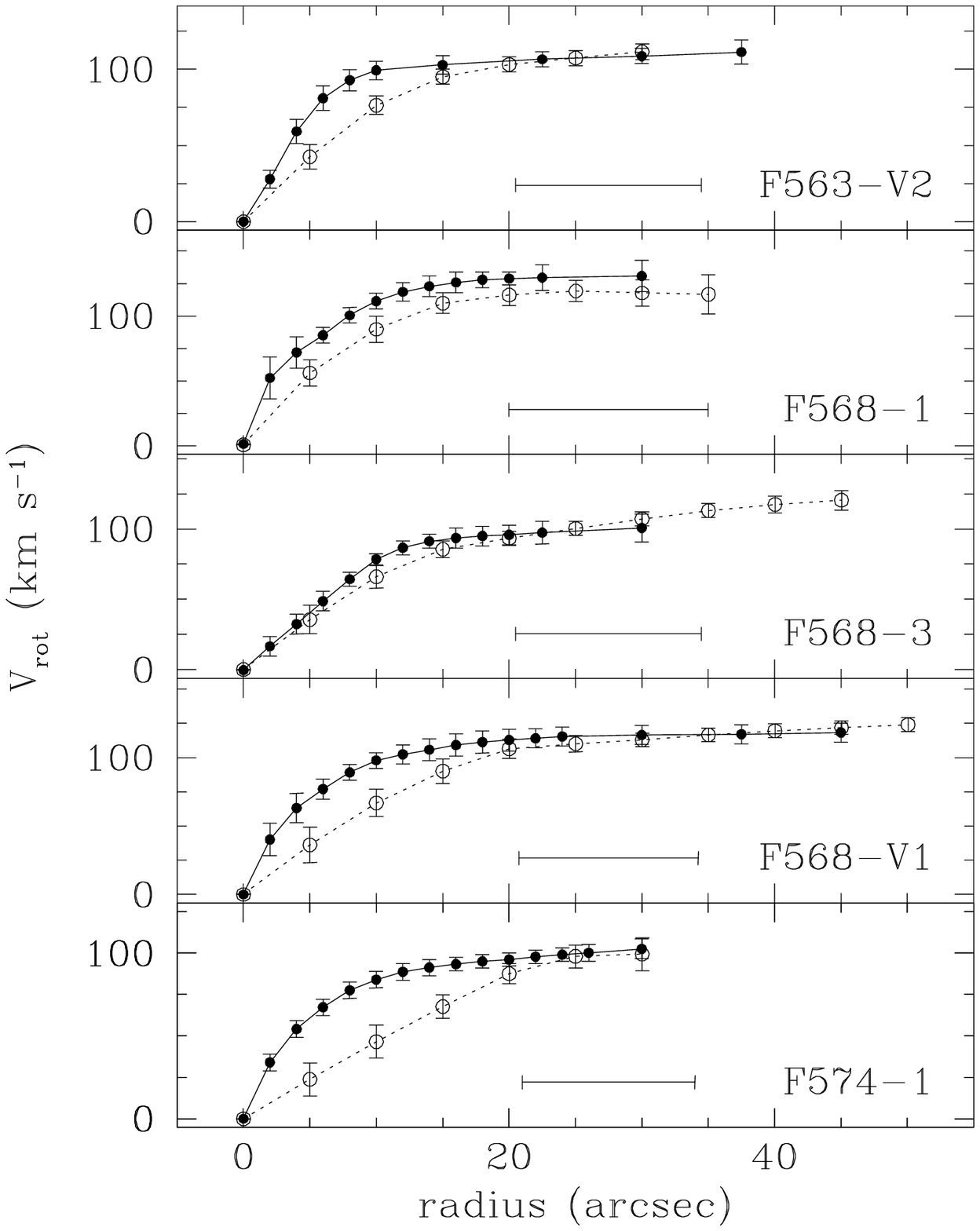}
\figcaption{The high resolution H$\alpha$ rotation curves
  (filled circles, solid lines) and the \HI\ rotation curves from BMH 
  (open circles, dotted lines). The horizontal bar shows the FWHM beam size
  of the \HI\ observations.}
\vskip6pt
\end{center}}

\noindent ciently bright to allow to determine the position of the center along
the slit with an accuracy of better than $1''$.

\begin{figure*}[ht] 
\epsfxsize=18.5cm   
\epsfbox{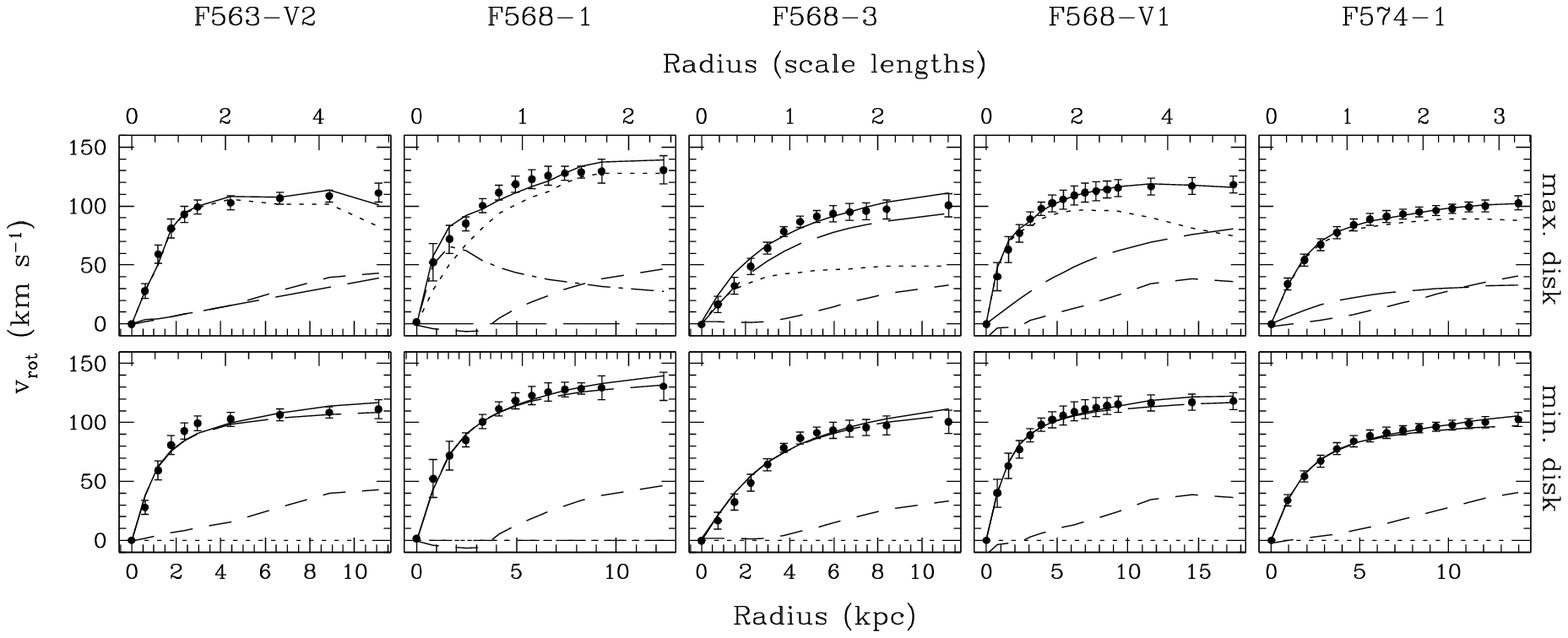}
\vspace{-0.4cm}
\figcaption{ Mass models fitted to the high resolution rotation
  curves. The top panels give maximum disk fits, the bottom panels
  give fits with a stellar mass-to-light ratio of zero. The dotted
  line represents the contribution of the stellar disk to the rotation
  curve, the dashed line the contribution of the gas, the long-dashed
  line represents the dark halo, and the full line represented the
  total model rotation curve. For F568-1, the dot-dashed line
  represents the contribution of the central component to the rotation
  curve.\label{figure3}}
\vskip-0.0cm
\end{figure*}

To obtain a larger radial coverage of the rotation curves, the H$\alpha$
data were combined with the \HI\ data presented in BMH.  To this end,
the derived H$\alpha$ velocities were plotted on the \HI\
position-velocity diagrams.  Both sets of data were found to agree well
with each other if the effects of beam smearing on the \HI\ data are
taken into account (cf.\ Swaters 1999).  Therefore, we have used the
\HI\ data to determine rotation velocities beyond radii where we found
H$\alpha$ emission.  Note that in most cases the \HI\ extends only
little beyond the H$\alpha$.  Next, the rotation curves derived for the
approaching and the receding sides were combined.  The H$\alpha$ points
were sampled every $2''$, the \HI\ points every $7.5''$ (approximately
two points per beam).  The errors on the rotation velocities were
estimated from the differences between the two sides and the
uncertainties in the derived velocities.  We will refer to these
rotation curves as the high resolution rotation curves (HRC). 

The derived HRCs are shown in Fig.~\ref{figure2} together with the \HI\
rotation curves presented in BM.  Probably because BM did not correct
for beam smearing, the \HI\ rotation curves systematically underestimate
the inner slopes of the rotation curves, especially for F568-V1 and
F574-1.  Both these galaxies have a central depression in the \HI\
distribution, as can be seen in the \HI\ maps presented in BMH.  The
spatial smearing of \HI\ from larger radii into the observed central
depression leads to an apparent solid body-like rotation curve, in
particular for the highly inclined galaxy F574-1. 

\section{Mass modeling}

For the mass modeling presented here we have used the same parameters
for the thickness of the gaseous and stellar disks as BM have done.  The
stellar disk was assumed to have a vertical sech$^2$ distribution, with
a scale height $z_0=h/6$.  $R$-band light profiles, presented in de
Blok, van der Hulst, \& Bothun (1995) and BMH, were used to calculate
the contribution of the stellar disk to the rotation curve.  The \HI\
was assumed to reside in an infinitely thin disk.  The only difference
with the mass models presented in BM is that we have decomposed the
light profile of F568-1 into a disk and a central component, and fitted
these components to the rotation curve separately.  In the other
galaxies no significant central component is present.  For the dark
matter component a pseudo-isothermal halo was used, following BM, which
has a rotation curve given by: \begin{displaymath}
v^2_\mathrm{halo}(r)=4\pi G\rho_0 r_c^2 \left[ 1-{{r_c}\over{r}} \arctan
\left( {{r}\over{r_c}} \right) \right], \end{displaymath} where $r_c$ is
the halo core radius and $\rho_0$ is the central density.

One of the major uncertainties in fitting mass models to rotation
curves, in absence of an independent measurement of the stellar
mass-to-light ratio \mlstar, is the uncertainty in the contribution of
the stellar disk to the rotation curve.  However, lower and upper limits
on \mlstar, and hence on the dark matter content, can be obtained by
assuming that the contribution of the stellar disk to the rotation curve
is either minimal or maximal.

In the maximum disk mass models, the contribution of the stellar disk
to the rotation curve is scaled to explain most of the inner parts of
the rotation curve.  The resulting rotation curve fits are shown in
the top panel of Fig.~\ref{figure3}.  What immediately strikes the eye
is that, in contrast with the findings of BM, the\break

\def\nodata{$\cdot\cdot\cdot$}
\begin{center}
\vskip-17pt
{\sc Table 2\\ \smallskip Mass model parameters \smallskip}
\footnotesize
\setlength{\tabcolsep}{6pt}
\begin{tabular}{lccccccc}  
\tableline\tableline\tskip 
&& \multicolumn{3}{c}{max.\ disk} &&
\multicolumn{2}{c}{no disk} \nl
\noalign{\vskip-3pt}
&& \multicolumn{3}{c}{\hrulefill} &&
\multicolumn{2}{c}{\hrulefill} \nl  
\hfil LSBC && \mlstar & $r_c$ & $\rho_0$ && $r_c$ & $\rho_0$ \nl
\hfil name && M$_\odot/\mathrm{L}_{R,\odot}$ & kpc &$\mathrm{M}_\odot
\mathrm{pc}^{-3}$ && kpc & $\mathrm{M}_\odot \mathrm{pc}^{-3}$ \nl
\tskip \tableline \tskip
F563-V2$^a$&& 5.4 & \nodata & \nodata & &0.94& 0.283 \nl
F568-1$^b$& & 17.2& \nodata & \nodata & & 1.5& 0.181 \nl
F568-3    & & 1.5 & 3.0& 0.027 & & 2.5&  0.48 \nl
F568-V1   & & 9.3 & 6.7& 0.005 & & 1.2& 0.188 \nl
F574-1    & & 3.7 & 3.4& 0.003 & & 1.5&  0.92 \nl
\tableline
\end{tabular}
\end{center} 
\vspace{-6pt}
\centerline{\vbox{\hsize=8cm \footnotesize $^a$ No $R$-band data are
    available, mass modeling is based $B$-band data.}}
\centerline{\vbox{\hsize=8cm \footnotesize $^b$ For F568-1 a central
    component was fitted separately, which has
    $\Upsilon_{\!\!*,\mathrm{bulge}}=14.4$ in the maximum disk fit.}}
\vskip12pt

\noindent inner parts of the rotation curves can be explained almost entirely
by the contribution of the stellar disk in all of these LSB galaxies,
with the exception of F568-3.  The dark halo parameters (see
Table~\ref{tabfits}) are ill-defined for most galaxies in our sample,
because most of the HRCs do not extend to large radii.  Nonetheless,
it is clear that in these maximum disk fits the contribution of the
dark halo will only become important outside the optical disk, as is
also the case for HSB galaxies.

The required stellar mass-to-light ratios for the maximum disk fits
(listed in Table~\ref{tabfits}), may be high, up to 17 in the
$R$-band.  Most of these are well outside the range of what current
population synthesis models predict (e.g., Worthey 1994).  If these
high values of \mlstar\ are to be explained solely by a stellar
population, the stellar content and the processes of star formation in
LSB galaxies need to be very different from those in HSB galaxies.
Alternatively, these high mass-to-light ratios may indicate the
presence of an additional baryonic component that is associated with
the disk, as has been suggested by e.g.\ Pfenniger, Combes, \&
Martinet (1994).  On the other hand, the fact that stellar disk can be
scaled to explain the observed rotation curve may simply reflect the
possibility that the luminous and dark mass have a similar
distribution within the optical galaxy.

\vbox{\begin{center}
\epsfxsize=\hsize   
\epsfbox{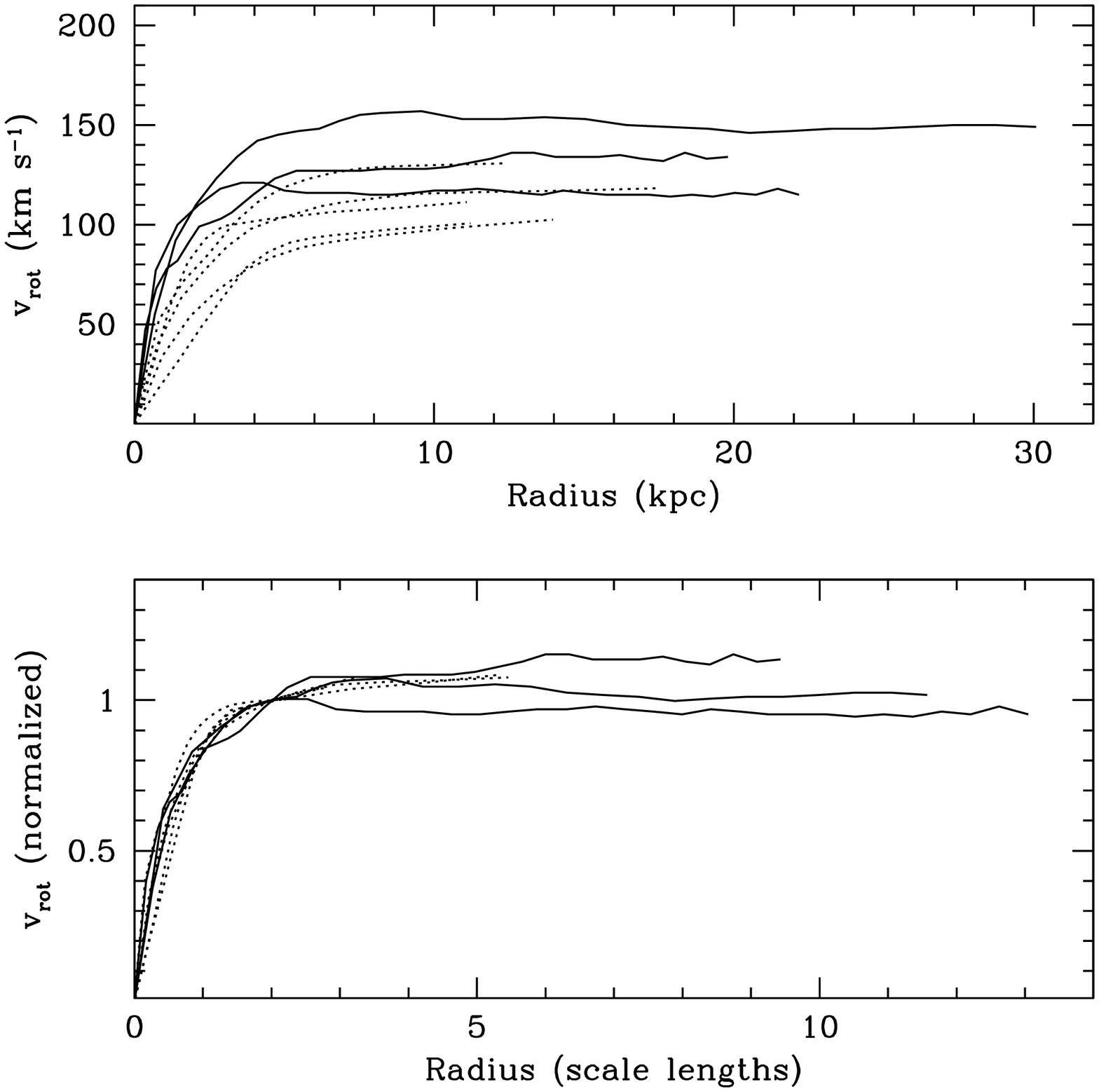}
\figcaption{Rotation curves of the five LSB galaxies
  (dotted lines) compared to the rotation curves of three typical
  late-type HSB spiral galaxies from Begeman (1987): NGC~2403
  ($M_B=-19.3$), NGC~3198 ($M_B=-19.4$) and NGC~6503 ($M_B=-18.7$).
  In the top panel, the rotation curves are expressed in kpc, in the
  bottom panel the rotation curves are scaled with their scale
  lengths, and normalized with the rotation velocity at two disk scale
  lengths.\label{figure4}}
\end{center}}

The other extreme for the contribution of the stellar disk to the
rotation curve is to assume that its contribution is negligible.  The
dark halo parameters for this minimum disk fit are listed in
Table~\ref{tabfits}.  High central densities of dark matter and small
core radii are required to explain the observed steep rise in the HRCs. 
From Fig.~\ref{figure3} it is clear that the minimum disk mass models
fit the rotation curves equally well as the maximum disk models.  In
fact, a good fit can be obtained with any mass-to-light ratio lower than
the maximum disk mass-to-light ratio, demonstrating that the degeneracy
that exists in the mass modeling for HSB galaxies (e.g.\ van Albada et
al.\ 1985) also exists for LSB galaxies.  Irrespective of the
contribution of the stellar disk to the rotation curve, the similarity
between the shapes of the observed rotation curves and those of the
stellar disks implies that the total mass density and the luminous
mass density are coupled within the region of the optical disk.

\section{Discussion}

The HRCs derived for the LSB galaxies rise steeply in the inner parts,
and reach a flat part beyond about two disk scale lengths, as is found
for HSB galaxies.  In Fig.~\ref{figure4}, the rotation curves for LSB
galaxies (dotted lines) are compared with those of three typical
late-type HSB galaxies from Begeman (1987), NGC~2403, NGC~3198 and
NGC~6503.  All the galaxies in Fig.~\ref{figure4} have no bulges, or
only weak ones.  In the top panel of Fig.~\ref{figure4} the radii are
given in kpc.  In these units, the rotation curves of LSB galaxies rise
more slowly than those of HSB galaxies, indicating that these galaxies
not only have lower central surface brightnesses, but also lower central
mass densities, as was found by de Blok \& McGaugh (1996) as well.

A different picture emerges in the bottom panel of Fig.~\ref{figure4},
in which the rotation curves are scaled by their optical disk scale
lengths and normalized to the velocity at two disk scale lengths.  The
normalization probably does not introduce systematic effects because the
maximum difference in absolute magnitude between the galaxies in
Fig.~\ref{figure4} is only 1.5 mag.  With radii expressed in disk scale
lengths, the rotation curves of the LSB galaxies presented here and
those of HSB galaxies have almost identical shapes.  This is consistent
with the concept of a `universal rotation curve' (Persic, Salucci, \&
Stel 1996, see also Rubin et al.\ 1985). 

The similarity between the LSB and HSB rotation curve shapes suggests
that rotation curve shapes are linked to the distribution of light in
the stellar disks, independent of central disk surface brightness.
Although such a link is most easily understood if the stellar disks are
close to maximal, independent of surface brightness, the required high
mass-to-light ratios seem to favor a picture in which LSB galaxies are
dominated by dark matter within the optical disk, and HSB galaxies more
by the stellar disk.  The relative contribution of the stellar disk to
the rotation curve may change continuously from LSB to HSB, or perhaps
LSB and HSB galaxies constitute discrete galaxy families, as has been
suggested by Tully \& Verheijen (1997).

\acknowledgments

We thank Roelof Bottema for valuable discussions, and Erwin de Blok for
making available the \HI\ data and for useful comments.  RS thanks IPAC
for its hospitality during his visits which were funded in part by a
grant to BFM as part of the NASA Long-Term Space Astrophysics Program.

\end{document}